\documentclass[aps,twocolumn,showkeys]{revtex4}
\usepackage{graphicx}       
\usepackage{dcolumn}        
\usepackage{bm}             
\usepackage{float}
\usepackage{amsmath, amsthm}
\usepackage{array}
\usepackage{notoccite}
\usepackage[sort&compress]{natbib} 

\begin{document}
\def\linenumberfont{\normalfont\tiny\sffamily}

\markboth{K.Chang, R.A. Murdick, Z. Tao, T-R.T. Han, C-Y. Ruan}{Ultrafast electron diffractive voltammetry: General formalism and applications}

%
%

\title{Ultrafast electron diffractive voltammetry: General formalism and applications}

\author{\footnotesize Kiseok Chang, Ryan A. Murdick, Zhensheng Tao, Tzong-Ru T. Han, Chong-Yu Ruan\footnote{Corresponding author: ruan@pa.msu.edu}}

\address{Department of Physics and Astronomy, Michigan State University,\\
East Lansing, Michigan 48824-2320,
USA}



\begin{abstract}
We present a general formalism of ultrafast diffractive voltammetry
approach as a contact-free tool to investigate the ultrafast surface
charge dynamics in nanostructured interfaces.  As case studies,
the photoinduced surface charging processes in oxidized silicon surface
and the hot electron dynamics in nanoparticle-decorated interface are
examined based on the diffractive voltammetry framework. We identify
that the charge redistribution processes appear on the surface, sub-surface, and vacuum
levels when driven by intense femtosecond laser pulses. To elucidate the voltammetry
contribution from different sources, we perform controlled experiments
using shadow imaging techniques and N-particle simulations to aid the investigation of the photovoltage dynamics in the presence of photoemission.
We show that voltammetry contribution associated with photoemission has a long decay tail and
plays a more visible role in the nanosecond timescale, whereas the ultrafast voltammetry are
dominated by local charge transfer, such as surface charging and molecular charge transport at nanostructured interfaces.
We also discuss the general applicability of the diffractive voltammetry
as an integral part of quantitative ultrafast electron diffraction
methodology in researching different types of interfaces having
distinctive surface diffraction and boundary conditions.
\end{abstract}

\keywords{interfacial charge transfer; diffractive voltammetry; ultrafast electron diffraction; femtosecond laser.}

\maketitle

\section{Introduction}

Electron transfer is a primary process responsible for energy transduction
 at interfaces, especially as the relevant length scale approaches 1 nm.\cite{Adams2003,Gratzel2003,AVIRAM1974,Avouris2007,Bezryadin1997,Haruta1997}
Transfer of an electron from the donor to the acceptor sites across an interface involves the coupling of the occupied electronic states of the former and the unoccupied ones of the latter,\cite{MARCUS1985,Chen1993,Datta_Book} and through photoexcitation, the hot carrier generation allows wider access to densities of states for the charge transfer, thereby yielding higher electrical conductance. Through molecular engineering of the interface\cite{Ashkenasy2002} and the implementation of nanostructured materials, \cite{Hu1999,Alivisatos1996,Kong2000,Shipway2000,Kongkanand2008,Leschkies2007,Luther2008} higher efficiencies are being realized in directed carrier transport through interfaces,\cite{Adams2003} with the expected fundamental $RC$ time reaching $\sim$1 ps. Using femtosecond (fs) laser pulses to excite carriers,  the elementary processes, such as interfacial hot carrier transport and relaxations through interacting with phonons and impurities near the surface that are essential to the efficiency of photovoltaic and photocurrent generation, can be investigated.\cite{Schmittrink1987,Benisty1991,Anderson2005,MillerBook} In addition to the general interest in nanoscale charge transfer phenomena mentioned above, generation of photoelectric field near surface from fs laser pulse is a subject of interests on its own, \cite{Petek2005,Fuss2000,vanDriel1999} mainly due to the novel aspects of fs laser which allows high pulsed laser photofield and the ultrafast electronic excitation that generates non-equilibrium electron distribution. The hot electrons thus generated provide separation of electron and holes, or interfacial charge transfer through enhanced photoconductivity, and at a high peak intensity multiphoton-induced processes, such as direct injection of interfacial molecular states, or even photoemission that produces external charge distribution can occur.
 Characterizing the microscopic interfacial charge transfer (forward and backward) beyond the initial steps of charge separation is central to the development of efficient solar energy transduction devices,\cite{Kongkanand2008,Anderson2005,Robel2006} nanoelectronics,\cite{Tian1998,Aswal2006,Chen1999}, reactive surface photochemistry, \cite{Haruta1997,Dulub2007,MISEWICH1992,Henderson2003} and nanostructure fabrication.\cite{Miyamoto2007,RamanPRL2008}
Recently, the ultrafast electron diffraction technique is shown to be able to investigate the charge transfer dynamics at
interface due to the sensitivity of the probing electrons to the transient electric field
distribution,\cite{RamanPRL2008,Murdick2008,ZewailAPL2010} causing a modification of the surface diffraction pattern,\cite{Spence1994,Spence1983,Spence2004} which is loosely characterized as the `refraction effect'.\cite{RefarctionNote} The methodology of measuring the interfacial photovoltage following photo-driven charge transfer via monitoring changes in the Bragg diffracted electron beams can be characterized as a ultrafast electron diffractive voltammetry (UEDV). In this paper, we extend on the previous work of electron diffractive voltammetry\cite{Murdick2008,MurdickThesis} with an aim to identify the different constituents of the measured transient
surface voltage (TSV) and discuss their respective roles in Coulomb refraction. We also develop a general formalism  that can quantitatively
describe these phenomena based on surface diffraction features, and provide examples to demonstrate the feasibility in different types of systems, including surface charging, interfacial charge transfer, and photoionization.

\section{Origins of transient photoinduced surface voltage}

\begin{figure}[th]
\includegraphics[width=1.0\columnwidth]{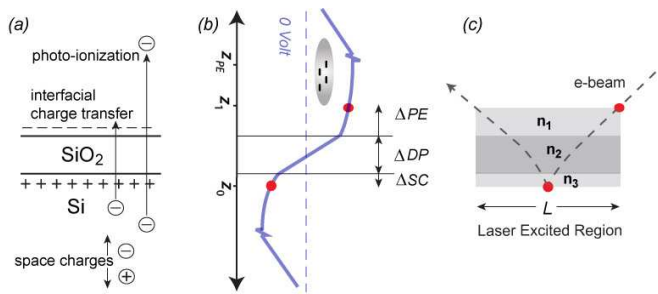}
\vspace*{8pt}
\caption{(Color online) Transient photoinduced charge redistribution near surafce. (a) The three mechanisms of photoexcitation that cause redistribution of charges at the surface, bulk, and vacuum levels near $Si/SiO_2$ surface. (b) Transient surface potential diagram caused by various photoinduced charge redistribution. (c) The refraction of the electron beam in each field region can be modeled by an index of refraction with $n=\sqrt{(\Delta V+V_0)/V_0}$, where eV$_0$ is the electron beam energy. \label{f1}}
\end{figure}

The photoinduced transient surface charge redistribution can appear in the sub-surface level (carrier separation, electronic excitation),
at interfaces (charge transfer), and above the surface (photoemission), as exemplified in Fig.~\ref{f1},
and the magnitude of surface voltage ($V_s$) can be described generally from the sum of these components:
\begin{equation}
V_s = \int^{z_1}_{z_0} E_z(z) dz,
\label{TSVInt}
\end{equation}
\noindent where z$_1$ is the position at which the probing electron beam enters the field region and  z$_0$ is the position
of the diffractively probed region. Effect of the these charge redistribution channels are categorized into respective photovoltages.
Firstly, on the bulk level, the inner potential change $\Delta$IP resulted from the adjustment of valence electronic distribution can cause an abrupt refraction shift at the interface, whereas the adjustment/creation of a space charge region defines a transient voltage $\Delta$SC within the dielectric screening length.
Secondly, on the surface level, the photoinduced interfacial charge transfer over a thin insulating barrier can create a dipole field region, which defines
a voltage component $\Delta$DP on a nm length scale. Thirdly, photoemission can occur into the vacuum region, particularly for high intensity excitations.   Subsequently a subsurface charge dynamics is induced to screen the field associated with photoelectron from penetrating into the bulk materials.
Together, they create a near surface field imparting a potential difference $\Delta$PE on a time-dependent
length-scale defined by the recovery of the photoelectrons to the surface.  These four
different mechanisms of surface voltage generation have characteristic time and length scales, their influences on the probing electron beam ultimately vary with incidence and exit angles and the interfacial structure. Generally, these photovoltages are linearly superposable, which allows them to be treated independently, and the overall surface voltage can be expressed as:
\begin{equation}
V_s = \Delta IP+\Delta SC+\Delta DP+\Delta PE.
\end{equation}

As a prototype example, photoinduced charge redistribution at Si/SiO$_2$ interface (Fig.~\ref{f1}) is examined here,
in which the voltammetry is contributed significantly from $\Delta$DP as the probing electron beam fully penetrate the top SiO$_2$ layer, whereas it has only a short penetration depth ($\approx$ 1 nm) into the Si underlayer.
Since the screening length in semiconductor is relatively large($\approx$ 1 $\mu$m), the short penetration of the probing electron beam picks up only a small fraction of the voltage drop along the top SC region, whereas the surface dipole voltage across the SiO$_2$ layer is fully sampled.  Thus at a scenario where interfacial charge transfer occurs,  $\Delta$DP can dominate over $\Delta$SC. Meanwhile, $\Delta$IP is generally small if phase transition is not involved. The contribution of $\Delta$PE is more difficult to assess. In the case of Si, the hot carriers are created with a high transient temperature, which can induce thermionic emission, and under an strong photofield from an intense laser irradiation the multiphoton photoemission is also possible,\cite{vanDriel2007} leading to a nonnegligible photoelectron contribution to the overall photovoltage. Nonetheless, due to the very different length scales involved in interfacial charge transfer and photoemission,
we expect the dynamics to be rather different, which will be investigated with controlled experiments.

\section{Surface diffraction and rocking map characterization}
\label{RockingCurve}

\begin{figure}[th]
\includegraphics[width=1.0\columnwidth]{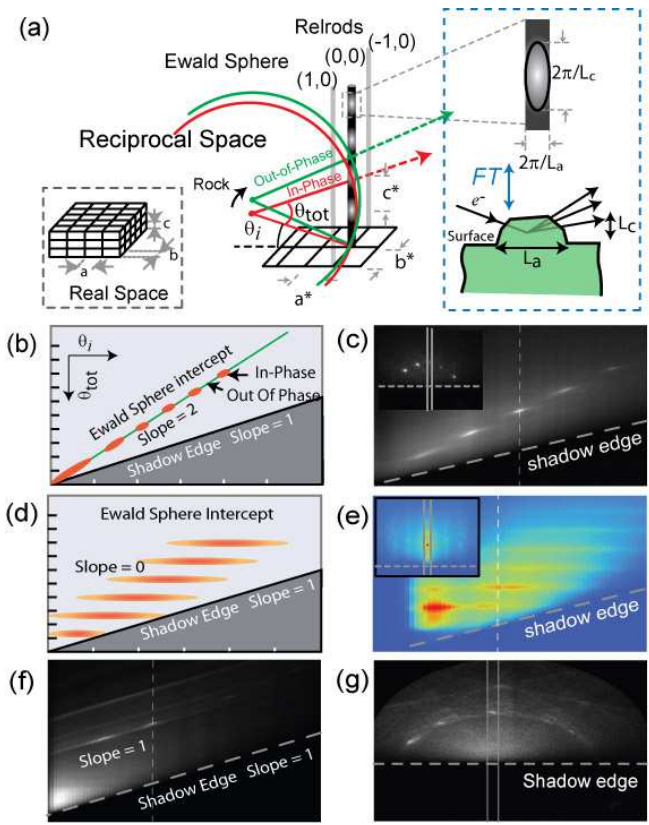}
\vspace*{8pt}
\caption{(Color online) Surface electron diffraction pattern in different conditions. (a) Ewald sphere construction in the grazing incidence angle geometry. By tilting (rocking) the angle of incidence between the electron beam and the sample, the Ewald sphere intercepts the reciprocal lattice rods (relrods) at different height. The in-phase condition is satisfied when the intercept is at the reciprocal lattice node.  The inset shows the reciprocal node structure, which is effectively determined from a Fourier Transform (FT) of the crystalline region in the sample defined by its persistence lengths.  (b) Expected rocking map of a smooth, pristine surface in RHEED. (c) Experimental rocking map taken from a smooth Si/SiO$_2$ surface. The dashed line shows where the diffraction pattern in the inset is taken. (d) Expected rocking map of a nanostructured surface. (e) Experimental rocking map pattern taken from a highly oriented pyrolytic graphite (HOPG) surface.  (f) Experimental rocking map taken from a $Si/SiO_2$ surface sample along a Kikuchi-enhanced diffraction peak. (g) Diffraction pattern of the $Si/SiO_2$ surface, showing visible Kikuchi pattern. \label{f2}}
\end{figure}

Since diffractive voltammetry employs diffracted beams, it is necessary to link the photo-induced distortion of diffraction pattern
with the surface voltage generation. First, we examine the formalism of
electron diffraction from different types of surface, which has been a source of
confusion to properly understand the ultrafast surface electron diffraction process and a central topic to elucidate for deducing $V_s$. Fig.~\ref{f2}(a) describes
the production of the diffraction pattern from a grazing incident electron beam. The
Ewald sphere is constructed to predict the diffraction pattern based on the intercept regions between the Ewald sphere and the reciprocal lattice network.
This methodology is founded on a kinematic (Fourier) theory and can be extended to understand nanoscale diffraction, in which the size of the reciprocal lattice nodes, as depicted in the inset of Fig.~\ref{f2}(a), is determined by the persistent length\cite{PersistenceLengthNote} of the lattice probed by the electrons.  For a long-range-ordered smooth surface,
the in-plane persistent length is very large (L$_a$ in the inset of Fig.~\ref{f2}(a)) as compared to the penetration depth of the electron (L$_c$), producing very thin reciprocal rods (relrods). In the limit of L$_c$=0 (single layer), the reciprocal lattice becomes two-dimensional (2D) array of relrods, and the diffracted beam is defined by the intercept
between the Ewald sphere and the relrod network, rendering circular diffraction patterns,  generally
described as the Laue zones in reflective high-energy electron diffraction(RHEED).  For nanostructured surfaces, $L_c\approx L_a$, so relrod widens significantly, and the diffraction pattern can deviate significantly from circular Laue patterns, and observing more than one diffraction peaks along a single relrod is possible.

To examine the relrod structure, we use rocking map characterization, which is conducted by rocking the sample plane against electron incidence, so Ewald sphere intercept rolls along the relrod. The rocking map is constructed by slicing a reflection stripe showing a relrod normal to the shadow edge in the diffraction image and stitching these stripes together as a function of incidence angle ($\theta_i$).
A diagonal line with slope (a) equal to 2 in the rocking map exposes the relrod structure in terms of $\theta_{tot}$ vs. $\theta_i$,
as depicted in Fig.~\ref{f2}(b). The reciprocal node, which is a Fourier transform of a probed crystalline
region defined by the persistence lengths of the samples as depicted in the inset of Fig.~\ref{f2}(a), can be examined from the out-of-phase to in-phase conditions in the rocking map.
Near $\theta_i$=0, the relrod structure is continuous. As $\theta_i$ increases the relrod
becomes spotty, due to the increase of persistent length (L$_c$) with the increasing electron penetration depth
as a function of $\theta_i$. This trend is evidenced in an experimental rocking
map produced from a relatively flat Si surface, as shown in Fig.~\ref{f2}(c). For the sake of clarity in discussion, we will refer to RHEED pattern only when dealing with a smooth surface
that produces sharp circular Laue patterns, which is especially useful for monitoring the layer-by-layer growth in molecular beam epitaxy.\cite{AppLEED}
In so speaking, RHEED experiment is not well suited for studying structural dynamics study as neither does the position of the RHEED
peak indicate the respective position of the reciprocal lattice node, nor does RHEED intensity directly inform lattice fluctuations, such as Debye Waller factor. Only through the inspection of the rocking map can the reciprocal lattice be exposed for structural investigation, nonetheless, such experiments are tedious to perform for dynamics study.\cite{RuanPNAS2004}

Fortunately, more informative results can be obtained for nanostructured surfaces and interfaces where the diffraction mechanism differs
from `RHEED'. In fact, typical ultrafast electron crystallography (UEC) studies,\cite{RuanScience2004,Baum2007,Gedik2007,Carbone2008,RamanPRL2008} rely on transmitted surface diffraction features produced with the grazing incidence electrons to determine structural dynamics. When the transverse persistent length ($L_a$ in the inset of Fig.~\ref{f2}(a)) is short, such as steps and nanostructure-decorated surfaces, the widened relrods can extend several periods of the interferences (Bragg reflections), taking advantage of the high energy electron having a large Ewald sphere radius ($\approx$90$\AA^{-1}$ at 30 keV) for extensive overlap with reciprocal lattice. As a result, the slope of the in-phase diffraction in the rocking map changes from 2 to 0, as it is now possible to penetrate the samples and produce transmission patterns.  A simulated rocking map (Fig.~\ref{f2}(d)) shows this trend, which is verified by a study of highly oriented pyrolytic graphite (HOPG),\cite{RamanPRL2008}
shown in Fig.~\ref{f2}(e).  This type of features differ from `RHEED' in that the
transmitted diffraction spots carry the symmetry of the lattice and can be used for structural determination. With UEC operated in such circumstances, the intensity
of transmitted Bragg peaks have been used to monitor the integrity of the lattice structure,
including laser-induced thermal fluctuations and phase transition,\cite{RamanPRL2008}
and have been exploited to investigate the surface-supported nanoparticles. \cite{NanoLetter2007,RamanPRL2011}
What's essential here for formulating the surface diffractive voltammetry is that
the diffraction condition under $a=0$ has: $\theta_i + \theta_o = \theta_{tot} =n \lambda/c$, where $\lambda$ is the electron wavelength, n is the diffraction order, and c is the lattice constant, can be used to formulate the diffracted beam trajectory under the presence of transient surface field, as described in Fig.~\ref{f3}. Details of the general formalism of UEDV for $a$=0 or 2 under different surface diffraction conditions will be described in detail in Sec.~\ref{SlabModel}.
We also like to point out it is generally difficult to know the circumstances of surface diffraction without examining rocking map. For example, when employing resonance diffraction peaks appearing along a Kikuchi line\cite{ZLWang} for UEDV, the surface diffraction must be characterized according to $a=1$, as shown in Fig.~\ref{f2}(f)(map) and (g) (Kikuchi pattern).  For this reason, it is central to identify the surface diffraction circumstance from the rocking map characterization before a proper interpretation of the data can be established.

\section{The general formalism of electron diffractive voltammetry}
\label{SlabModel}

\begin{figure}[th]
\includegraphics[width=1.0\columnwidth]{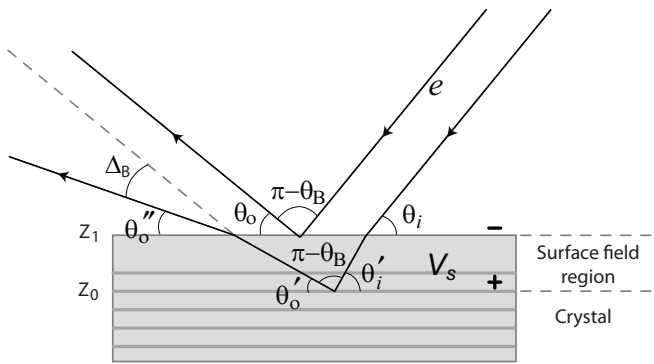}
\vspace*{8pt}
\caption{The idealized slab model for considering the transient surface voltage.
The top trajectory is the electron scattering from the crystal planes with the presence of a
surface field. The electron beam, incident at $\theta _i$, is Bragg scattered at $\theta _B$,
exiting the surface at $\theta _o$. Introducing an attractive surface potential $V_s$ will cause
the electron beam to be 'refracted' deeper into the crystal($\theta _i'$) and the same for the Bragg
diffracted beam that would ultimately exit the crystal at $\theta _o"$ with a net shift $\Delta _B$
relative to $\theta _o$. \label{f3}}
\end{figure}

We derive the general formalism for describing the TSV-induced distortion of
the diffraction pattern under different surface diffraction circumstances. We firstly
generalize the problem in a simplified infinite long slab geometry, as depicted in Fig.~\ref{f3},
where a field region exists near the surface, caused by a photoinduced
redistribution of charges.  We consider the 'refraction' effect separately for
the incident and outgoing beams.   As the electron beam enters the slab, the
incidence angle is changed from $\theta_i$ into $\theta_i'$ due to the refraction effect imposed by the surface
field region. A similar refraction effect occurs as the diffracted beams cross the
same region to reach the detector screen, which changes the exit angle from $\theta_o'$
in the diffraction region to $\theta_o"$ as the diffracted beam leaves the field region.
The degree of change depends on the strength of field
integrated along the incident and exit paths. Due to the grazing incidence geometry the change
in $\theta$ is dominated by the field normal to the surface, we can relate the
change in $\theta$ to $V_s$ based on momentum- energy relationship along z direction
for the incoming beam:

\begin{equation}
{p_{z_1}^i}^2 - {p_{z_0}^i}^2 = 2m_eeV_s, \label{ECons1}
\end{equation}

\noindent where $p_{z_1}^i$ and $p_{z_0}^i$ are the momenta of the incident beam projected
along $z$ at $z_0$ and $z_1$. Expressed in terms of angle $\theta$, Eq.~(\ref{ECons1})
can be rewritten as:

\begin{equation}
\tan^2 \theta_i' = \tan^2 \theta_i + \frac{\chi}{\cos^2 \theta_i}\label{ECons2},
\end{equation}

\noindent where $\chi=V_s/V_o$, by utilizing tan$\theta_i=p_{z_1}^i/p_x$,
tan$\theta_i'=p_{z_0}^i/p_x$, and $eV_0$ is the beam energy prior
entering the field region. Similarly for the outgoing beams,
we have:

\begin{equation}
\tan^2 \theta_o' = \tan^2 \theta_o'' + \frac{\chi}{\cos^2 \theta_o''} \label{ECons3}.
\end{equation}

Since the electric field integration is linear, different components of the surface potential can be superposed on each other, thus the details of $V_s$ composition is not important here. The voltammetry is established when $V_s$ can be deduced as a function of the observable $\Delta_B$, which is defined as the angular shift of the diffracted
beam ($\Delta_B\equiv\theta_o"-\theta_o$). The derivation of $\chi(\Delta_B)$ requires the knowledge of surface diffraction.
To make the voltammetry generally applicable to different type of interfaces, we consider all scenarios discussed in Fig.~\ref{f2}
by relating $\theta_o$ and $\theta_i$ with
\begin{equation}
\Delta \theta_o=(a-1)\Delta \theta_i \label{ECons4},
\end{equation}

\noindent where $a$ is the slope along the in-phase diffracted beams in the
rocking map. For example, $a=2$ belongs to the case of RHEED (Fig.~\ref{f2}(b)), $a=0$
is associated with the transmitted Bragg diffraction (Fig.~\ref{f2}(d)), and $a=1$ can be attributed to Kikuchi diffraction (Fig.~\ref{f2}(f)). Following the notation in Fig.~\ref{f3}, at
the diffracted region:

\begin{equation}
\theta_o'=(a-1)(\theta_i'-\theta_i)+\theta_o, \label{ECons5}
\end{equation}

\noindent which allows us to rewrite $\tan\theta_o'$ in Eq.~(\ref{ECons3}), which we define as $D$,
 in terms of $\theta_i$ and $\theta_o$:

\begin{equation}
\tan \theta_o' = \frac{\tan [\theta_o+(1-a)\theta_i]-\tan [(1-a)\theta_i']}
{1+\tan [\theta_o+(1-a)\theta_i] \tan [(1-a)\theta_i']} \equiv D,
\label{ECons6}
\end{equation}

\noindent where $\theta_i' = \tan^{-1}(\sqrt{\frac{\sin^2 \theta_i+ \chi}{1-\sin^2 \theta_i}})$
according to Eq.~(\ref{ECons2}). From Eq.~(\ref{ECons3}) at given $V_s$, $\theta_i$ and $\theta_o$:

\begin{equation}
\theta_o'' = \sin^{-1} \sqrt{\frac{D^2-\chi}{1+D^2}}.
\label{ECons7}
\end{equation}

\noindent To get $V_s$-induced angular shift in the diffraction pattern:

\begin{equation}
\Delta_B = \sin^{-1} \sqrt{\frac{D^2-\chi}{1+D^2}} - \theta_o.
\label{ECons8}
\end{equation}

\begin{figure}[th]
\includegraphics[width=1.0\columnwidth]{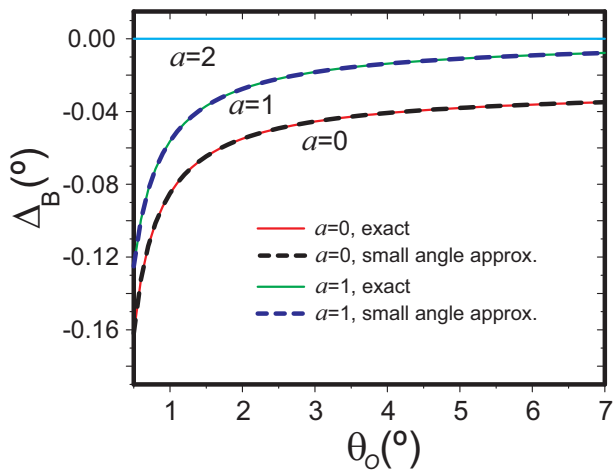}
\vspace*{8pt}
\caption{(Color Online) The refraction-induced shift($\Delta_B$) for diffraction peak located at $\theta_o$ at $V_s$=1 Volt calculated for difference surface diffraction condition characterized by $a=0,1,2$ (see Fig.~\ref{f2}). The solid lines are exact solution from voltammetry formalism. The dashed lines are calculated employing small angle approximation (see text). The incidence angle ($\theta_i$) is set at 2.01$^\circ$.  \label{f4}}
\end{figure}

Since $D$ is a function of $\chi$, it is difficult to deduce $\chi(\Delta_B, \theta_i, \theta_o)$ directly from Eq.~(\ref{ECons8}). Small angle approximation allows inverting Eq.~(\ref{ECons8}) to obtain $\chi(\Delta_B)$ for different $a$, which is presented in the Appendix.
One salient feature of the refraction-induced shift is that $\Delta_B$ increases as $\theta_o$
decreases. This is easily seen in Fig.~\ref{f4}, where $\Delta_B$ is calculated for $a=0,1,2$ at $V_s=1V$, following the exact solution based on Eq.~(\ref{ECons8}).
We note that the corresponding change in $\Delta_B$ at $a=0$ is nearly twice the value at $a=1$, whereas at $a=2$, $\Delta_B$ remains to be 0 for all
$\theta_o$. We also calculate $\Delta_B$ using small angle solutions (Eqs.~(\ref{AEqn7}) \& (\ref{AEqn9})) in the Appendix. The difference between the two is barely noticeable. These results show that voltammetry is best performed using nanostructured materials where the transmitted Bragg diffraction is possible (i.e. $a=0$), whereas UEDV would be impossible under strictly RHEED condition. Nonetheless, in real circumstances even for a relatively flat surface, $a$ is usually not exactly equal to 2, as shown in
Fig.~\ref{f2}(c). Deviation from $a=2$ results in a small, but non-negligible sensitivity to $V_s$.  Generally,
the angular dependence of $V_s$-induced shift is opposite to the structure-related one,
as the Fourier relationship: $d\theta/\theta \sim -dr/r$ demands
that if only the structural change is present $\Delta_B$ would increase as $\theta_o$ increases. In contrast, the refraction-induced shift responds to $V_s$ oppositely, resulting in non-uniform cancellation of structure-induced shift. This nonreciprocal feature is the basis
of a Fourier Phasing method,\cite{MM2008} used to correct the Vs-induced distortion in
the diffraction pattern in order to accurately assess the structural dynamics.

\section{Ultrafast electron diffractive voltammetry experiment}

\begin{figure}[th]
\includegraphics[width=1.0\columnwidth]{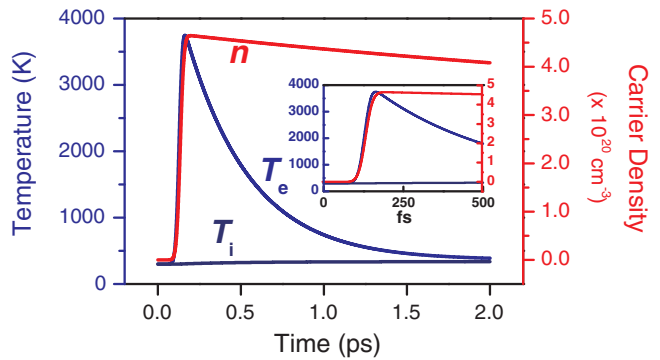}
\vspace*{8pt}
\caption{(Color Online) Non-thermal Boltzmann transport coupled two-temperature model calculation results of
electronic/ionic temperature ($T_e, T_i$) and carrier density induced by 800 nm femtosecond
laser in silicon. The laser fluence and pulse width are $65 mJ/cm^2$ and 45 fs,
and the silicon sample thickness is $50 \mu m$. Inset: Closeup view of the first 500 fs. \label{TTM}}
\end{figure}

To demonstrate the methodology for UEDV, we conduct an experiment using a Si (111) substrate
with a thin ($\sim 2nm$) insulating $SiO_2$ layer prepared via modified RCA cleaning.\cite{Kern1970}
We employ a Ti:Sapphire laser (800nm) with photon energy of 1.55 eV, which is above
the indirect bandgap (1.11 eV) of Si but below the bandgap of SiO$_2$ (8.9 eV), to excite
the carriers within the Si substrate. Using the Boltzmann transport coupled
two-temperature model,\cite{MurdickThesis,9,Carpene2006} we can estimate the transient carrier density, and the quasi-equilibrium electronic and lattice temperatures
in the electron probed region.\cite{MurdickThesis} The coupling hierarchy for energy transduction is that the excitation photon first heats up the carriers through
electronic relaxation in the sub-ps timescale, whereas the
lattice temperature rises mainly through electron-phonon coupling on the ps timescale.
In parallel, the energy transport takes place from excited surface into
the bulk, driven by the temperature gradient, mediated by the carriers and the phonons.
Because for Si the penetration depth ($l$) of the infrared laser is significantly longer
($1 \mu m$)\cite{Jellison1982} than that of the electron beam ($\sim 2 nm$,
incident angle $6.8^o$), induction periods ($\tau_i$) for the temperatures
to decay at the surface exist and can be estimated based on $\tau_i=l^2/D$,
where D is the diffusivity of the electrons and the phonons.
We find that the the $\tau_i$ for electron ($\sim$ 700 ps) is significantly longer than electron-phonon
coupling time ($\sim$ 5 ps), thus allowing the stored photon energy to be maintained within the photoexcited
region to yield a lattice temperature rise on 5-10 ps
timescale.  Nonetheless, due to the long penetration depth and large disparity in the
electronic and lattice heat capacities, even at a relatively high fluence of 65 $mJ/cm^2$, the lattice temperature
rise is very small compared to that of the carriers having kinetic energy on par with the
excitation energy, as revealed from the two-temperature model calculation shown
in Fig.~\ref{TTM}. The lattice temperature rises by only 40 K, yielding a thermal expansion
of the lattice at most 0.011\% ($~3.4 \times 10^{-4} \AA$).  In addition to the transient
high temperature, the carrier concentration can increase by more than 3 orders
of magnitude in the surface excited region from the intrinsic level,\cite{Chen2005} thereby creating a favorable condition for
studying hot electron driven interfacial charge transfer across the $SiO_2$ layer.

\begin{figure}[th]
\includegraphics[width=1.0\columnwidth]{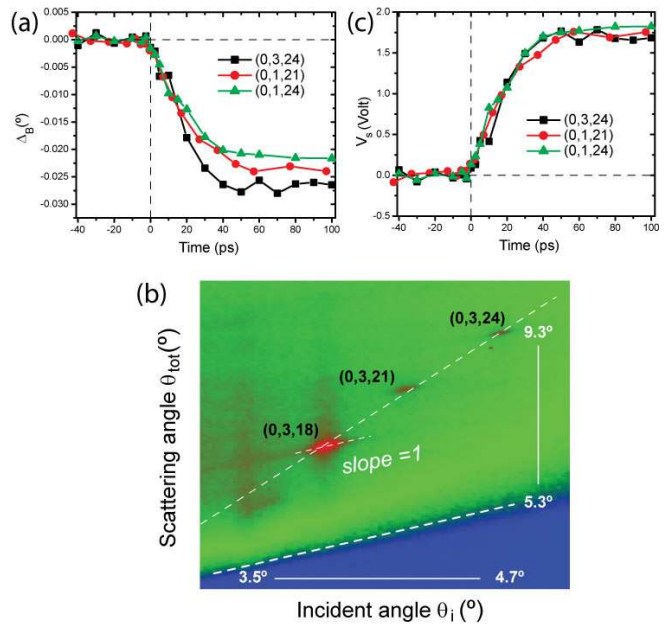}
\vspace*{8pt}
\caption{(Color Online) Trasient voltammetry from three diffracted beams from Si/SiO$_2$ interface. (a) The angular shift of (0,3,24), (0,1,21), and (0,1,24) beams excited at F=65mJ/cm$^2$. (b) The rocking map characterization of (0,3)-relrod, showing a surface diffraction condition $a=1$. (c) The photovoltage deduced from (0,3,24), (0,1,21), and (0,1,24) beams based on Eq.~(\ref{ECons8}) using $a=1$. \label{ThreeBeams}}
\end{figure}

The signature of refraction-induced shift in the diffraction pattern lies in its angular dependence.
We choose to investigate (0,3,24), (0,1,21), (0,1,24) on the (0,3) and (0,1) relrods with $N$=7 and 8 (crystallographic notation for
 diffraction order is multiplied by a factor of 3 due to the ABC layering of the Si(111) surface), corresponding to
$\theta_i$ of 6.24$^{\circ}$, 4.15$^{\circ}$, $\&$ 4.70$^{\circ}$, and $\theta_o$ of 3.50$^{\circ}$,
 4.02$^{\circ}$, $\&$ 4.63$^{\circ}$ respectively.  The excitation fluence is fixed at 65 mJ/cm$^2$. Under photo-illumination, the transient movement of the three diffracted beams, depicted in Fig.~\ref{ThreeBeams}(a), indeed exhibits nonreciprocal signatures as described previously, i.e. the higher order Bragg peak shifts less than the lower order one,
 which is characteristic of the TSV-induced effect.\cite{Murdick2008}  Closer examination of the shifts shows that nonreciprocity applies only to $\theta_o$, but not to $\theta_i$, as the maximally shifted beam is (0,3,24), which has a $\theta_i$=6.24$^{\circ}$ larger than the rest, whereas its corresponding $\theta_o$ is 3.50$^{\circ}$, which is smaller than the rest. This angular dependence is confirmed by the rocking map analysis for the relrods exhibiting $a=1$ near the in-phase diffraction, as shown in Fig.~\ref{ThreeBeams}(b)) for (0,3) relrod. By applying $a=1$ in Eq.~(\ref{ECons8}), we deduce $V_s$ for the three diffracted beams and render consistent TSV curves, independent of $\theta_i$. Having established the validity of voltammetry measurement, we  investigate the sources of the voltammetry in this study, which can be associated with suface/subsurface charge dynamics and/or photo-ionization creating photoelectrons.

\begin{figure}[th]
\includegraphics[width=1.0\columnwidth]{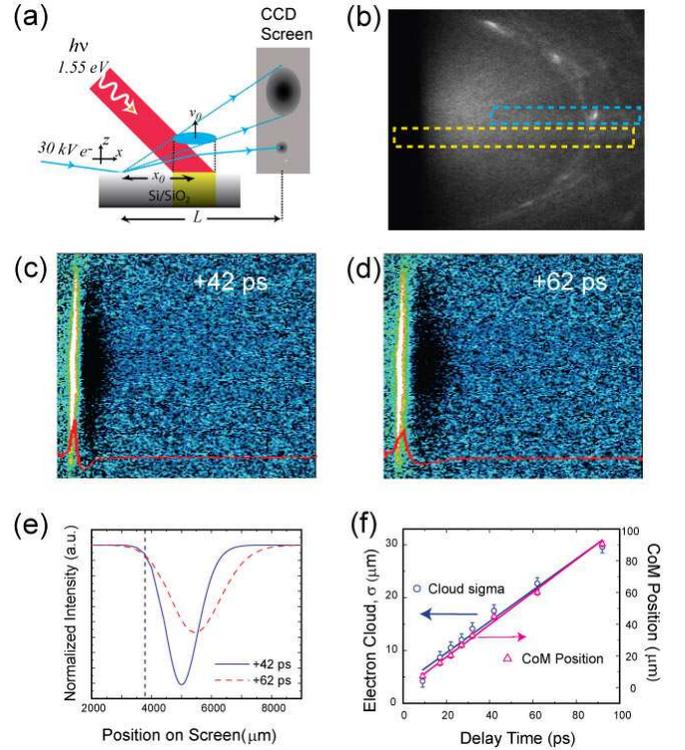}
\vspace*{8pt}
\caption{(Color Online) Shadow imaging experiment to characterize the properties of photoemission. (a) Schematic experiment setup of the experiment, in which the incident electron beam  is displaced by $x_0$ from the photoinduced region by 800nm pump laser. The surface scattered electrons form a shadow image of the electron cloud on the CCD screen as they are scattered away from the collective field associated with photoelectrons. In parallel, the surface diffracted beam
experiences the electric field associated with photoemitted electron cloud, and deflects according to its location relative to the cloud. (b) The diffraction pattern from Si/SiO$_2$ surface is shiwn with the stripe regions selected for extracting the shadow image evolution (yellow) and the diffracted beam reflection (cyan). (c) \& (d) show the snap-shot shadow images of the photoemitted electron cloud at different time delays. (e) The respective Gaussian fitting of the shadow images. (f) Results extracted from fitting the shadow image of the photoemitted electron cloud, showing the evolution of the CoM position and the cloud width. \label{ShadowSetup}}
\end{figure}

To isolate the contribution of free electrons produced by photo-ionization in the vacuum region,
$i.e.~ \Delta PE$, we conduct a controlled study in which the photoelectron
dynamics and the surface photovoltage are characterized simultaneously. This is achieved by using ultrafast electron shadow
projection imaging approach, which has been reported previously for studying
photoemission from an HOPG surface.\cite{RamanAPL2009} The advantage of electron
projection imaging technique is that it can be
implemented \emph{in situ} with the voltammetry experiment by simply displacing the
electron beam from the pump-probe overlap position by a distance ($x_0$) (Fig.~\ref{ShadowSetup}(a)), thereby
investigating the photoelectron dynamics under the same excitation condition as the voltammetry experiment.
In addition, the diffracted beams, which are also visible in the shadow images,
are affected only by photoemitted electrons above the surface generated by the pump
laser, but not affected by the subsurface fields probed in the voltammetry geometry, thus
establishing a clean way to evaluate the effect of $\Delta PE$ in voltammetry.

In principle all the relevant information pertaining to photoemission
for creating the transient near-surface field can be obtained by the projection
imaging study.  Fig.~\ref{ShadowSetup}(c \& d) show two selected snap-shots of the shadow images
of the photoemitted electron cloud obtained at t=42ps and 62ps under a laser fluence of 65 $mJ/cm^2$.
The initial lateral dimension of the electron cloud is determined by our pump laser incident at $45^o$ to the surface normal.
As a result, the laser footprint is elliptical with $\sigma_x=330 \mu m$ and $\sigma_y=233 \mu m$, which are determined by the cross-correlation response by scanning the probe
beam across the laser illuminated region.\cite{MM2008}  The projection
distance (source-to-camera) employed in this study is 16.5 cm, and the offset
distance $x_0=2.23 mm$, giving a magnification factor $\sim 74$.  The linescan of the shadow images (integrated vertically along the yellow stripe depicted in Fig.~\ref{ShadowSetup}(b)) contains the respective temporal evolvement of Gaussian-like electron cloud together with a near-surface build-up (lines colored in red in Fig.~\ref{ShadowSetup}(c \& d), and from fitting
the linescans\cite{RamanAPL2009} (Fig.~\ref{ShadowSetup}(e)) at different times the temporal evolution of cloud width ($\sigma_z$) and
center-of-mass position ($z_{CoM}$) can be determined, as depicted in Fig.~\ref{ShadowSetup}(f)
for $t=0 \sim 100~ps$. From these temporal profile changes, we observe a linear increase of position and width, and extract an electron cloud expansion velocity $v_{\sigma_z}=
0.336 \mu m/ps$ and an initial CoM velocity $v_{CoM}=1.02 \mu m/ps$.

The creation of shadow images can be understood based on scattering of
the probing electrons from the collective field established by the
photoelectrons:

\begin{equation}
E_{PE}=\frac{1}{4\pi \epsilon_0} \sum^N_{i=1} \frac{e}{(r-r_i)^2}.
\label{Nparticle1}
\end{equation}

\begin{figure}[th]
\includegraphics[width=1.0\columnwidth]{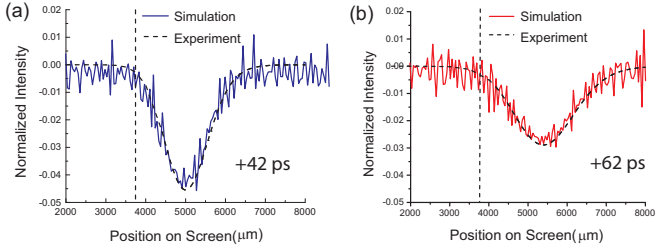}
\vspace*{8pt}
\caption{(Color Online) N-particle shadow projection imaging simulation at two different time delays. \label{f8}}
\end{figure}

\noindent The deflection from the collective field reduces the numbers of originally forward-going
electrons reaching the CCD camera, thus effectively creating a shadow of the
electron cloud. This process can be directly simulated by an N-particle
simulation\cite{Tao2012} employing Monte Carlo sampling of a 3D Gaussian electron distribution, which is
parameterized based on $\sigma$s obtained from fitting the shadow images. We then send rays of electrons representing the probing electron beam
across the 3D electron cloud whose collective field is calculated first by summing the pair-wise fields from individual electrons within the cloud. To speed up the calculation, we establish a mean-field model to match the results calculated from on the multi-particle calculation based on the impact parameter to the electron cloud.\cite{Tao2012} We note that the deflection caused by the collective field is linear with respect to the electron density in the regime of interest here, which warrants the usefulness of the mean-field approach. To simulate the shadow formation, $10^7$ electron rays are used along the line of sight to establish the shadow line scans and the electron counts on the CCD screen are calculated with and without intervening by the 3D electron cloud. The shadow profiles are constructed by dividing the electron counts along the line scans with the rays being intersected by the 3D could in the path and the line scans without intersection and compared with the experimental results. Fig.~\ref{f8} shows the comparison at two different delay times at t=42 and 62 ps (solid lines are N-particle simulation results and dashed lines are the fitted gaussian profiles obtained from experimental shadow images). The agreement between the N-particle simulations and the shadow imaging results are excellent. Since here the depth of the shadow is proportional to the photoemitted electron density, the agreement of between the experiment and simulation not only indicates the robustness of the shadow imaging technique in profiling the photoemission, but also offering a measurement of the photoelectron density created by the photo-illumination.

\begin{figure}[th]
\includegraphics[width=1.0\columnwidth]{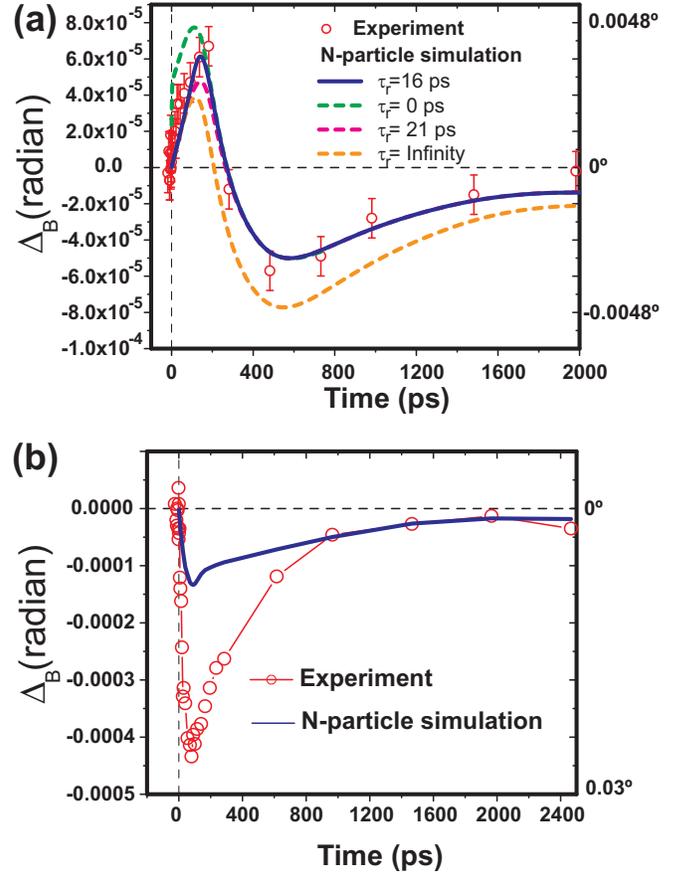}
\vspace*{8pt}
\caption{(Color Online) Experiments to characterize photoelectron dynamics and surface photovoltage performed at F=65 mJ/cm$^2$. (a) Data (symbols, colored in red) show the deflection of a selected diffracted beam by the electric field associated with photoelectrons and the image charges on the surface acquired in the shadow imaging experiment setup. N-particle simulations with surface dielectric relaxation times ($\tau_r$) ranging from 0, 16, 21 ps, and $\infty$ are used to fit the data. (b) The voltammetry results (symbols, colored in red) obtained from the same diffracted beam, but at the overlapped voltammetry geometry. An N-particle simulation to estimate the refraction contribution associated with photoemission is shown (solid line, colored in blue) for comparison.    \label{f9}}
\end{figure}

An independent approach to deduce the photo-electron density is through a single-beam deflection experiment across the photo-emitted electron cloud.\cite{ParkAPL2009} Importantly, the 3D cloud geometry established by the shadow imaging technique can be confirmed by the deflection of a diffracted beam from Si(111) surface diffraction, as it traverse through the collective field associated with photoelectrons. Single-beam deflection experiment has the advantage of being highly sensitive to the field and so is applicable even at very long times (ns) when the electron cloud become very diffusive to monitor by the shadow imaging approach. To extend the field characterization to longer time, we analyze the Si-111 beam deflection data contained in the diffraction images obtained from shadow imaging experiment. The analysis is on a Kikuchi diffraction enhanced peak (with $\theta_i$=2.01$^\circ$ \& $\theta_o$=5.76$^\circ$) along the central stripe region circled by the dashed line in cyan in Fig.~\ref{ShadowSetup}(b)). Fig.~\ref{f9}(a) shows the temporal evolution of the angular shift. The up and down swings of the beam can be associated with the beam crossing from above and under the 3D electron
cloud.\cite{ParkAPL2009}  Importantly, the electron density required in correctly simulating the shadow profile can now be directly confirmed by simulating the
specific electron trajectory using an N-particle calculation as described earlier. Furthermore, the absolute downward shift of the diffracted beam is also affected by the counter image force associated with the image charges that are created on the surface responding to the photoemission, which is also modeled numerically as described below.

\section{Near surface field induced by photoemission}

To fully simulate the diffracted beam deflection trajectory, which extends to 2 ns,
we need to know the projectile motion of the 3D electron cloud and the
corresponding image charge dynamics that provide additional field component.
To comply with the rate and the magnitude of beam deflection, a metal-like dielectric response with a very large $\epsilon$ at early times due to the
excessive amount of charges initially built up on surface is considered.  $\epsilon$ decays
exponentially to the equilibrium value of 3.9 for SiO$_2$,\cite{MeyerBook} and can be
described by  $\epsilon_{SiO_2}$=3.9 + A $e^{-t/\tau_r}$.  The field model for image charges is
included with the dielectric relaxation process and is described by:\cite{JacksonBook}

\begin{equation}
E_{Img}=-\frac{1}{4\pi \epsilon_0}\frac{\epsilon_{SiO_2}-1}{\epsilon_{SiO_2}+1}
\sum^N_{i=1} \frac{e}{(r-r_i)^2}.
\label{Nparticle2}
\end{equation}

\noindent The time-dependent angular shift of diffracted beam is calculated using:

\begin{equation}
\Delta\theta=\frac{\int E dl}{2V_0\cos^2\theta},
\label{Nparticle3}
\end{equation}

\noindent
where $V_0$=30kV and E in the path integral contains contributions from
photoelectrons (Eq.~(\ref{Nparticle1})) and image charges (Eq.~(\ref{Nparticle2})). The integration takes place over 3$\times$FWHM across the Gaussian cloud.  Previously, an analytical model\cite{ParkAPL2009} and N-particle simulation\cite{ZewailAPL2010} of the transient field associated with photoelectron cloud and image charges have been implemented to account for deflection of a probing electron beam. The dielectric response of the surface has not been exclusively included.   The necessity in incorporating the surface dielectric relaxation to account for the change in $\epsilon$ is evident
from comparing the models with different dielectric relaxation times and the experimental data, which are shown in Fig.~\ref{f9}(a).
We find A=$10^{4}$ and $\tau_r$=16 ps provide a reasonable agreement to the experimental data.  To comply with the deflection data at long times, the knowledge of the photoelectron
cloud beyond the initial linear trajectory is needed. The return rate of the photoelectrons to the surface is determined by the strength of the image force, which is gradually weakened as the expansion of the photoelectron cloud into the surface will lead to cancellation of the image charges even before the CoM trajectory reaches the turning
point and weakens the image force. We apply an additional 3rd order
term to account for this effect. The deflection of the diffracted beam can be calculated self-consistently by
varying the coefficients of the 2nd and 3rd order polynomials to fit the deflection
data.   We note that the fitting is based on a fixed initial condition (electron density and CoM and expansion velocity) determined by shadow imaging, but
the results from fitting deflection data extend our knowledge of the surface field development beyond
the timescale (0-150 ps) obtainable from the shadow imaging, and serve as the
basis for estimating the long time behavior for the voltammetry measurement.

\section{Surface photovoltage}

Having obtained the near surface field associated with the photoelectrons
from the shadow imaging and diffracted beam deflection measurements in the offset
geometry, we can now evaluate the contribution associated with photoemission
in the voltammetry experiment conducted by shifting the beam from the offset geometry to the overlap geometry, as reported
in Fig.~\ref{f9}(b) (line and symbols, colored in red).  The diffracted beam used in the voltammetry experiment appears at the intercept of a Kikuchi line and a 2D reciprocal lattice rod (Fig.~\ref{f2}(g)). We characterize the diffraction being a two-step process, where the incident beam is first scattered randomly to form an isotropic source, and then scattered into surface Laue Zones.  This is consistent with $a=1$ observed in the $\theta_{tot}$ vs $\theta_i$
relationship obtained from the rocking map analysis presented in Fig.~\ref{f2}(f), implying that only the
refraction along the exit path contributes to $\Delta_B$, and so we can simplify the
generalized TSV formula accordingly, and deduce the photovoltage based on: $\chi=-\Delta_B(\Delta_B+2\theta_o)$ (see Appendix, Eq.~(\ref{TSVa1})).

\begin{figure}[th]
\includegraphics[width=1.0\columnwidth]{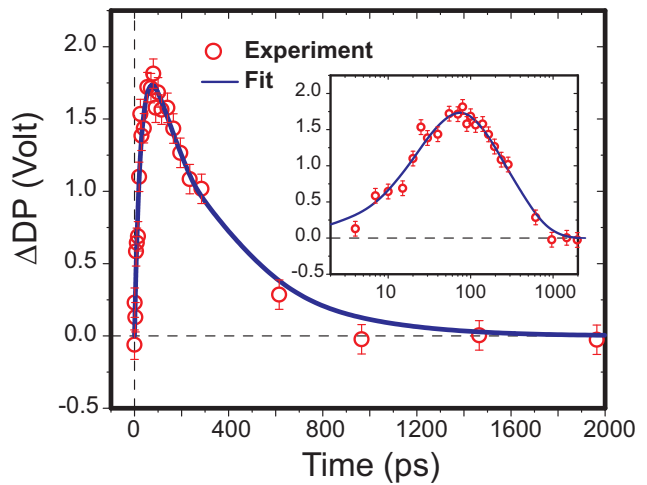}
\vspace*{8pt}
\caption{(Color Online) Transient surface voltage caused purely by $\Delta DP$. $\Delta PE$ contribution is subtracted
from total TSV, and the surface voltage is calculated by Eq.~(\ref{RCFit}). The surface voltage
is fitted by an RC charging and discharging model with $\tau _c$=30.84 (ps), and $\tau_d$=296.47 (ps).
Inset: Charging/discharging dynamics in a log time scale.\label{f10}}
\end{figure}

We evaluate the contribution $\Delta$PE in the overall photovoltage measurement by building on the knowledge of near surface fields induced by photo-emission characterized by the shadow imaging and deflection experiments.  The $\Delta_B$ associated with $\Delta$PE can be calculated by an N particle simulation of $\Delta_B$ along the exit beam path at $\theta_o$, similar to that in evaluating the deflection experiment, but under an overlap geometry used in the voltammetry experiment. Thus-simulated $\Delta_B$ is depicted in Fig.~\ref{f9}(b) (solid line, colored in blue) to represent the $\Delta$PE contribution, and compared to the overall $\Delta_B$ measured experimentally. Remarkably, the photoemission-associated
$\Delta_B$ matches very well with the long-time tail of the voltammetry measurement, but
contributes maximally about 25-30\% of the total angular shift at the early times. This indicates that the slow dynamics of TSV at the long time are controlled by the return of the photoemitted electrons to the surface, whereas interfacial charge transfer across the SiO$_2$ layer is more relevant at the short times.
By excluding the $\Delta$PE contribution from the overall $\Delta_B$, we deduce a V$_s$(t) relevant to the surface charging dynamics, as depicted in Fig.~\ref{f10}.

By fitting the $\Delta$PE subtracted V$_s$(t) with an effective RC-charging/discharging model:
\begin{equation}
V_s=V_{fit}(1-e^{-t/\tau_c})e^{-t/\tau_d},\,
\label{RCFit}
\end{equation}
\noindent
we determine RC time constants: $\tau_c$=30.84 ps, and $\tau_d$=296.47 ps respectively for surface charging and discharging.
We attribute the difference between the two to the change of hot electron photoconductivity
across the SiO$_2$ interface.  The photo-generated hot electrons facilitate the surface charging through access to the excited states, leading to a much shorter RC time than the discharging, which involves a cooler interface with a reduced photoconductivity, leaving the SiO$_2$ surface to stay charged for a longer period of time.
Surface charging processes have previously also been investigated by electric field induced second harmonic generation (EFISH),
\cite{LEE1967,CHEN1983,MIHAYCHUK1995,Aktsipetrov1996,Dadap1997,Baldelli2000} in which the sub-surface electric
field is deduced based on modeling the field-enhancement of optical second harmonic generation signal as well as photoemission.\cite{Marsi2000}
We find that the field strength $E$ $\sim$ 1-5 V/nm, obtained in our study of the Si/SiO$_2$ interface,
is very similar to what was found in EFISH studies under similar excitation conditions,\cite{Aktsipetrov1996,Dadap1997}
but because of a lower laser repetition rate is applied here (1 kHz in UEDV, compared to ~80 MHz in EFISH), cyclic residual charge accumulation from deep trap states\cite{Mihaychuk1999,Scheidt2008,Jun2004a,Tolk2007} is avoided, allowing the transient charging behavior to be resolved directly.

We compare the TSV results reported here for a smooth Si/SiO$_2$ interface with one reported previously for a step Si/SiO$_2$ interface. We find that the timescales in charging and discharging the interface in the two studies are similar, whereas the TSV induced in the smooth interface (maximum 1.7V at 65 mJ/cm$^2$) is smaller than the stepped interface (maximum 3V at 72 mJ/cm$^2$). By applying the shadow imaging technique under the same conditions (electron incidence/exit angles and laser fluences) as the voltammetry experiment, we are able to quantitatively identify the contribution of photoemission on the overall voltammetry result, where photoemission is mainly responsible for the slow decay, but does not contribute significantly to the short time dynamics, which clarifies the origin of the diffracted beam movement.\cite{ParkAPL2009} For cases where photoexcitation can cause significant structural changes, a correction on the refraction-induced shift in diffraction pattern is needed. We point to the first ultrafast electron crystallography investigation of Si(111) surface, which was performed using 266 nm laser pulse.\cite{RuanPNAS2004} Because of the much shorter laser penetration depth (4nm), the absorbed optical density is concentrated near the surface, propelling not only hot electron dynamics, but also surface structural changes.  From examining the reported time-dependent rocking curve (Fig.4(a) in the paper\cite{RuanPNAS2004}), the movement of the `in-phase' Bragg peak at small angle ($\sim 3.1^\circ$) is slightly larger than that of a peak at higher angle($\sim 3.9^\circ$), which is consistent with a surface refraction phenomenon being present. But the surface charging is not the full story, as the `surface' structural dynamics was also examined in the `out-of-phase' condition(Fig.4(b) in the paper\cite{RuanPNAS2004}), where the presence of multiple interference peaks is a signature of transmitted diffraction from surface nanostructure. Such a pattern was modeled using a slab model that identified the changes are from the top surface layer and the sub-surface (111) layers, contributing the contrasting movements of the two different peaks separated in $\approx 30 ps$. The surface dynamics could be mediated by the impulsive strain induced by ultrafast laser pulse heating and/or surface charges. Further controlled study is needed to clarify the nature of the surface dynamics on surface charging, photo-emission and the corresponding structural dynamics, which can be achieved using the methodologies provided here(see discussion in Sect.~\ref{SecFourierPhasing}).

\section{Diffractive Voltammetry beyond slab model}
\label{SecBeyondSlab}
%
\begin{figure}
\includegraphics[width=1.0\columnwidth]{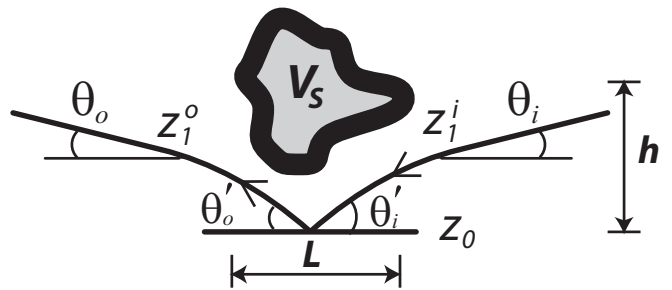}
\caption{General refraction geometry for grazing incident electron beam. The effective bending of the incident (exit) beam, caused by the local field associated with V$_s$ in the nanostructures, differs depending on the entry (exit) point $z_1^i$($z_1^o$). Since the relative change of the transverse momentum remains small ($V_s \ll V_0$), the slab model can be extended to treat the general refraction effect considering an angle- and position-dependent correction factor (see text).}
\label{GernalTSVGeo}
\end{figure}

While the general formalism deduced in Sec. \ref{SlabModel} is limited to a smooth interface in which the transient surface field is modeled to have a slab geometry, the basic concept of diffractive voltammetry is applicable to different types of interfaces beyond the slab geometry. For different geometries, the timescales of the charge redistribution can likely be deduced correctly from $\Delta_B$(t), whereas V$_s$(t) deduced using the slab-formalism is merely an effective parameter for TSV, which is subjective to corrections from the shape factor and the boundary conditions affecting the field integration at the nanointerface.  To treat TSV beyond an infinite slab model, we extend the TSV formalism described in Eq.~(\ref{ECons8}) to consider refraction shift with finite size and non-planar geometries. Under such circumstances, the location of $z_1$ in Eq.~(\ref{TSVInt}) needs not to be the same for incident and diffracted electron beams, and the effective $V_s$ that the electron beam experiences can be accounted for by applying the finite-size correction factors, $\Theta_i(\theta_i,\alpha)$ and $\Theta_o(\theta_o,\alpha)$ to separately describe the deviation of electron trajectory from the infinite slab model. This generalized picture, in which we can parameterize the finite interface structure with a nominal lateral length $L$ and vertical height $h$ with an aspect ratio parameter $\alpha\equiv h/L$, is depicted in Fig.~\ref{GernalTSVGeo}. The corresponding  $\Theta_i(\theta_i,\alpha)$ and $\Theta_o(\theta_o,\alpha)$ associated with the interface can be obtained numerically by ray tracing methods.
\par
We can then rewrite Eqs.~(\ref{ECons2})-(\ref{ECons8}) in terms of $\Theta_i(\theta_i,\alpha)$ and $\Theta_o(\theta_o,\alpha)$:

\begin{equation}
\tan^2 \theta_i' = \tan^2 \theta_i + \frac{     \Theta_i(\theta_i,\alpha) \chi   }
{\cos^2 \theta_i},
\label{EqnTan3}
\end{equation}

\begin{equation}
\tan^2 \theta_o' = \tan^2 \theta_o'' + \frac{     \Theta_o(\theta_o,\alpha) \chi   }
{\cos^2 \theta_o''},
\label{EqnTan4}
\end{equation}
and
\begin{equation}
\Delta_B=\sin^{-1}\sqrt{  \frac{D^2-\Theta_o(\theta_o,\alpha) \chi}{1+D^2}}-\theta_0 ,
\label{EqnDeltaB2}
\end{equation}
where
\begin{equation}
D = \frac{
\tan \left( \theta_i +\theta_o \right)
-\sqrt{\tan^2 \theta_i + \Theta_i(\theta_i,\alpha)\chi/\cos^2 \theta_i}
}
{
1+\tan(\theta_i+\theta_o)\sqrt{ \tan^2 \theta_i + \Theta_i (\theta_i,\alpha) \chi/ \cos^2 \theta_i  }
}.
\label{EqnD2}
\end{equation}
Here, $\chi \equiv V_s/V_0$. Applying small angle approximation allows the voltammetry formalism: $\chi(\Delta_B,\theta_i,\theta_o)$ to be established analytically, which is detailed in the Appendix. Below we give two examples to show how the correction terms can be implemented for dealing with more complex nanostructured interfaces beyond the infinite slab model.

\subsection{Correction for a finite slab geometry}
%
\begin{figure}
\includegraphics[width=1.0\columnwidth]{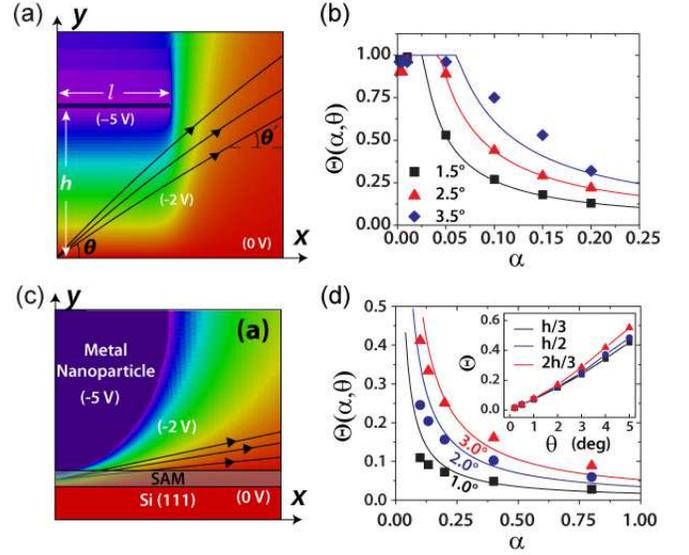}
\caption{ (Color Online) Corrections for the boundary condition associated with finite-size geometries. (a) The calculated potential for a slab of length $2l$, fixed at -5 V, a distance $h$ from a grounded base plane. (b)  The correction factor calculated for various aspect ratios, $\alpha = 2h/l$.  Each curve represents a different launch angle, $\theta_i$.  The solid lines are the predicted correction factors given by Eq.~(\ref{TSVCorr}). (c) The calculated potential for the case of a 20 nm metallic nanoparticle,
charged to -5 V, with a 1 nm thick dielectric layer ($\epsilon = 2.5$) on the Si substrate.
(d) The correction factor as a function of aspect ratio (SAM to nanoparticle) is calculated
by performing ray-tracing in the field distribution described in (a). The solid symbols are the simulated data; the lines are from Eq.~(\ref{TSVCorr}) using the definition of $\alpha$ from the text. Inset: The correction factor as a function of launch angle $\theta$ for three different launch positions along the SAM.}
\label{FigTSVCorrSlab}
\end{figure}

The simplest extension of an infinite slab formalism is to consider the edge effects in a finite slab geometry. Using aspect ratio $\alpha=2h/l$, where $h$ and $l$ are the height and width of the slab, as the probing electron beam enter and exit from the sides, the apparent voltage observed by the electron beams is reduced from $V_s$, which can be approximated by the correction factor:
\begin{equation}
\Theta(\theta,\alpha) = \left\{
\begin{array}{lr}
\theta/ \alpha \;\; \qquad  \mbox{if }\theta \; < \; \alpha, \\
1 \qquad \qquad  \mbox{if }\theta\; \ge \; \alpha.
\end{array}
\right.
\label{TSVCorr}
\end{equation}
\noindent To assess the validity of this approximation, we perform an electron ray tracing simulation \cite{FieldPrecision} with a setup shown in Fig.~\ref{FigTSVCorrSlab}(a) to calculate $\Theta (\theta, \alpha)$, which accounts for the fringe fields not included in Eq.~(\ref{TSVCorr}). We calculate a few instances of $\Theta (\theta, \alpha)$ as a function of $\alpha$ and show that the analytical expression described in Eq.~(\ref{TSVCorr}) is a fairly good approximation for $\theta <3.5^{\circ}$, as illustrated in Fig.~\ref{FigTSVCorrSlab}(b).  The finite-slab correction effectively suppresses the divergence of $\Delta_B$ as the diffracted electron beam approaches the shadow edge ($\theta_o$ approaches 0).

\subsection{Correction for a spherical nanoparticle decorated interface}
A frequently encountered molecular electronic interface involves replacing the top piece of the finite slab with a spherical nanoparticle, and has a molecular contact between the nanoparticle and the substrate. In order to model the TSV in this spherical interface, the correction factor $\Theta$ is parameterized as a function of angle $\theta_i$ ($\theta_o$). Ray tracing, as depicted in Fig.~\ref{FigTSVCorrSlab}(c), shows that the $\Theta (\theta,\alpha)$ (Fig.~\ref{FigTSVCorrSlab}(d)) determined for this spherical interface is amazingly similar to that of the flat finite slab interface (Fig.~\ref{FigTSVCorrSlab}(b)), if we re-define $\alpha=h/D$, where $D$ is the diameter of the nanoparticle and $h$ is the thickness of the molecular contact layer. For small angle diffraction ($\theta \le 2^{\circ}$), the dispersion in $\Theta$ due to position-dependent refraction effect can be ignored, as shown in the inset of Fig.~\ref{FigTSVCorrSlab}(d), calculated for the 20 nm nanoparticle interface.

\subsection{Charge transport in substrate-molecule-nanoparticle interface}
\begin{figure}
\includegraphics[width=1.0\columnwidth]{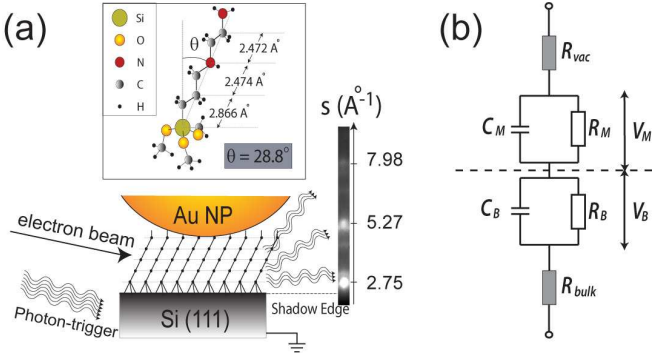}
\caption{(Color Online) (a) Schematic of an electron beam scattering from the ordered self-assembled monolayer (SAM) chain (inset), which serves as the linker that immobilizes the Au nanoparticles on the Si(111) substrate.  (b)  The equivalent circuit diagram to describe the charging/discharging dynamics in the semiconductor/SAM/nanoparticle interconnect geometry. Here `$M$' and `$B$' denote the resistance ($R$) and capacitance ($C$) associated with the SAM and the substrate, respectively.  The resistance of the bulk ($R_{bulk}$) and vacuum gap ($R_{vsc}=\infty$) is considerably larger than $R_M$ and $R_B$.
\label{FigTSVCorrNP}}
\end{figure}

The transient electron diffractive approach can be used to investigate silicon-molecule-gold nanoparticle interface, a prototypical system employed to study electron transport in molecaulr contacts.\cite{Bezryadin1997,Tian1998,Aswal2006,Chen1999,Forrest2004} This experimental scheme is shown in Fig.~\ref{FigTSVCorrNP}(a), where a self-assembled monolayer (SAM) is built on the hydroxylated Si substrate through silanization as the molecular contact, and covered by 20 nm gold nanoparticles\cite{Aswal2006,NanoLetter2007}.  The transient charging of the nanoparticle, caused by photoinduced charge transfer between the substrate and the nanoparticle through SAM, establishes a voltage determined by the charging energy of the nanoparticles w.r.t. the driving surface potential and the resistance of SAM, which can be conceptualized as an effective RC circuit as depicted in Fig.~\ref{FigTSVCorrNP}(b). The diffracted beams from SAM layer, order $N$=1 to 3, at $s$= 2.75, 5.27, 7.98 $\AA$ as shown in Fig.~\ref{FigTSVCorrNP}(a), are employed to investigate the charge transfer dynamics between the nanoparticles and the substrate. The capacitance can be calculated directly from the geometry of the interface using a finite element method.\cite{FieldPrecision} When C is known, the resistance of the SAM can be directly deduced by the RC time in the charging and discharging of the nanoparticle/SAM/Si interface. Due to that the interfacial molecular diffraction is also deflected by the imposing surface potential rise from the substrate, $V_s$ obtained from the overall $\Delta_B$ of the molecular diffraction consists of $V_M$, the voltage across the SAM, and $V_B$, which encompasses the surface charging and photoemission potential background.

\begin{figure}
\includegraphics[width=1.0\columnwidth]{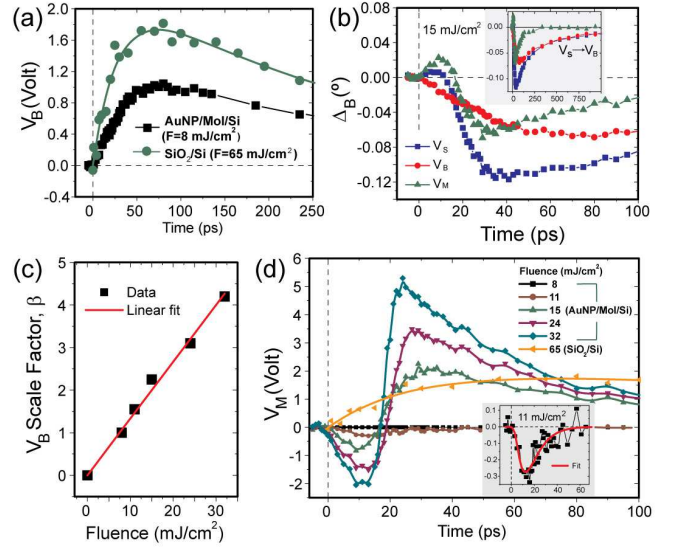}
\caption{(Color Online) Ultrafast transport at gold nanoparticle/SAM/silicon interface. (a) The surface voltage deduced based on voltammetry using the (001) diffracted beam (at $s=2.75$ \AA) from the SAM described in Fig.~\ref{FigTSVCorrNP} under F=8 mJ/cm$^2$, compared with the surface voltage deduced from a SiO$_2$/Si interface at F=65 mJ/cm$^2$, showing similar transient timescales. (b) The overall refraction shift determined by SAM diffracted beam (labeled $V_s$), the background (labeled $V_B$) obtained using $V_B$ from (a) and scaled to match the long time transient of $V_s$ (shown in the inset), and the molecular charge transport contribution, labeled $V_M$, obtained by subtracting $\Delta_B$(V$_B$) from $\Delta_B$(V$_s$). The data are obtained at F=15 mJ/cm$^2$. (c) The scale factor $\beta$ used in scaling $V_B$ to match $V_s$ in the long time for different fluences, showing a linear trend w.r.t fluence. (d) $V_M$ determined for different fluences, compared to the transient voltage obtained for SiO$_2$/Si at F=65 mJ/cm$^2$. Inset, the $V_M$ obtained at F=11 mJ/cm$^2$ is fit to a RC model with nearly equal time constants of 8$\pm$1 ps for charging/discharging.}
\label{FigMoleCharging}
\end{figure}

We first investigate $V_B$ by examining the low fleunce data (F$<$ 10 mJ/cm$^2$) where the driving force is insufficient to overcome the molecular potential barrier to charge up the nanoparticles. The voltammetry performed at F=8 mJ/cm$^2$ indeed shows a very similar transient to that obtained from bare SiO$_2$/Si interface (Fig.~\ref{f10}).  Comparison of the low fluence voltammetry result with that of SiO$_2$/Si is shown in Fig.~\ref{FigMoleCharging}(a), verifying that the low fluence $\Delta_B$ is solely from the surface charging potential of the Si surface, denoted here as $V_B$(8 mJ/cm$^2$). Increasing fluence beyond 10 mJ/cm$^2$ produces structures on the 10 ps timescale over the smooth $V_B$, showing first upward swing and then downward swing, exemplified in the $\Delta_B(V_B)$ subtracted $\Delta_B(V_s)$, or $\Delta_B(V_M)$, as exemplified in Fig.~\ref{FigMoleCharging}(b) for F=15 mJ/cm$^2$, which are characteristic of molecular charge transport.  We write
\begin{equation}
\Delta_{B}(V_{M})=\Delta_{B}(V_{s})-\Delta_{B}(V_{B}),
\label{AddingVM}
\end{equation}
\noindent , where $V_M$ is the voltage across the SAM. Here, the superposition principle is applied to $\Delta_B$s in Eq.~(\ref{AddingVM}), which is true when $\Delta_B \ll \theta_i,\theta_o$ (see Appendix for discussion). Since $V_M$ is driven by hot carriers in the nanoparticles and the Si substrate, $V_M$ should approaches 0 at a long time, or in other words, $V_s \rightarrow V_B$ as t $\rightarrow \infty$. We verify this by scaling up the $\Delta_B$(F=8 mJ/cm$^2$) to higher fluences according to $\Delta_B$(F)=$\beta$$\Delta_B$(F=8 mJ/cm$^2$) to compare $V_s$ and $V_B$ at long times, where $\beta$ is expect to be $\sim F/8$. Indeed, the long time transient of $V_s$ matches nearly perfectly with that of $V_B$ at F=15 mJ/cm$^2$, as shown in the inset of Fig.~\ref{FigMoleCharging}(b). We verify this for all the $V_s$ data at higher fluences. The values of $\beta$ obtained from comparing $V_s$ and $V_B$ are reported in Fig.~\ref{FigMoleCharging}(c), which shows a nearly linear relationship between $\beta$ and F, as expected.

With $V_B$ being well characterized, we are now poised to examine the photo-induced transport properties associated with SAM. Fig.~\ref{FigMoleCharging}(d) shows $V_M$ under different fluences. The $V_M$ is deduced first calculating $\Delta_B$($V_M$), according to Eq.~(\ref{AddingVM}), and then converting $\Delta_B$($V_M$) to $V_M$ by applying Eq.~(\ref{EqnChiLong}), in which $\Theta_i$=-0.0958 \& $\Theta_o$=-0.0234, calculated according to $\theta_i$=1.4$^\circ$ \& $\theta_o$=0.44$^\circ$ in our probing geometry. Examination of fluence-dependent $V_M$(t) reveals the following novel features compared to the surface charging dynamics of SiO$_2$/Si interface (also shown in Fig.~\ref{FigMoleCharging}(d) for comparison):

\noindent (1) The photovoltage sampled by SAM-diffracted beam is `negative' at first, indicating a net positive charge on the gold nanoparticles following photoemission. This shows that the initial photo-induced process is electron transfer from nanoparticles to Si surface. Since neither the carrier concentration nor the carrier temperature can be increased significantly for metallic nanoparticles by photoexcitation, as compared to Si substrate, the early electron migration to Si surface might suggest a laser-induced surface plasmon effect being present, excited in the gold nanoparticles that promotes a surface-field assisted or multi-photon internal ionization for nanoparticle charging.\cite{RamanPRL2011}

\noindent (2) A `reverse' charging process occurs as the fluence is increased beyond 11 mJ/cm$^2$, as evidenced in the surging of $V_s$ after 10 ps at F=15 mJ/cm$^2$, which reaches a positive value at $\sim$ 22 ps. The reversal of nanoparticle charging is driven by the overcharging of Si surface. The reversal time is short, within 5 ps at high fluence. The rapid reversal is indicative of a field-induced dielectric breakdown in SAM, which becomes a conductor as the interfacial field reaches a threshold value ($\approx$ 0.8 V/nm according to Fig.~\ref{FigMoleCharging}(d)). Following the breakdown, the relaxation of the stored electrons in the nanoparticles is much slower ($\sim$ 40 ps), indicating a recovery of the dielectric constant to the insulator status in SAM.

\noindent (3) The charging and reverse charging in nanoparticles are apparently not equal. The positive maximum $V_M$ (electron charging) is approximately twice as large as the negative maximum $V_M$ (hole charging) in Fig.\ref{FigMoleCharging}(d). This asymmetry in photoconductivity could be intrinsic to SAM, or it could be associated with the probing geometry in which the electron beam is more effectively deflected by Si surface than by nanoparticles due to a curved interface. In the latter case, the difference in maximum $V_M$ on the two polarities might not be associated with the difference in the stored charges within the nanoparticles. Further numerical evaluation of the asymmetry in the probing geometry is needed to clarify.

It is rather interesting to compare the molecular resistance deduced from this ultrafast voltammetry measurement with the steady-state resistance data,\cite{Sato1997} obtained by applying a bias voltage across the molecular interface, reporting an $R$ of 12.5 M$\Omega$ using 10 nm (thus a SAM area $\sim$ 4 times smaller) Au nanoparticle. We determine the nominal molecular resistance by obtaining the RC time constants from fitting $V_M$(t) recorded at F=11 mJ/cm$^2$, which is beneath the dielectric breakdown threshold. The fit using Eq.~(\ref{RCFit}), depicted in the inset of Fig.~\ref{FigMoleCharging}(d), shows a nearly equal $\tau_c$ and $\tau_d$ of 8$\pm$1 ps. Based on the effective RC model, we obtain a resistance $R_M=$ 2.74 M$\Omega$ using $C$=2.92$\times$10$^{-18}$ Farad, deduced from finite element modeling of the interface. The $R_M$ obtained using ultrafast voltammetry is 4 times smaller than the steady-state measurement for 10 nm particle, which shows the molecular resistivity obtained from two different methods are nearly identical.

\subsection{Diffractive voltammetry and ultrafast electron diffraction}
\label{SecFourierPhasing}
Evidenced from the general presence of photovoltage in photoexcited nanostructures, especially under intense laser excitation, in order to quantitatively determine the structural evolution, the refraction-induced shift in the diffraction pattern needs to be accounted for.  Fortunately, because of the `non-reciprocal' nature of the refraction shift, a self-consistent correction scheme can be implemented by applying a counter shift based on TSV-formalism to the diffraction spectrum. The correction aims to re-establish the Fourier relationship between the structure and diffraction pattern through iterative forward and backward Fourier transforms,\cite{RamanThesis,MM2008} which is termed Fourier Phasing (FP) scheme. The pre-requisite of an effective FP is that a sufficiently wide range of Fourier space encompassing several Bragg peaks is available for extract Fourier components, i.e. the pair-correlation function.\cite{MM2008}  This FP scheme has been successfully applied to extract the pair-correlation function for surface-supported nanoparticles\cite{MM2008} and graphite.\cite{RamanPRL2008} It has been shown that so long as a large portion of the Bragg spectrum is corrected, the structural changes, especially deduced based on the nearest neighbor distances, is robust, and the effective $V_s$ used to parameterize the refraction-shift can ne deduced reliably. Nonetheless, the absolute magnitude of TSV is subject to whether an exact TSV formalism has been identified, which is determined by the geometry and the surface diffraction condition.  This robustness arises from the inverse trend of the refraction shift in contrast to the structure-related shift and partially from the similarity between diffraction TSV formalisms, as discussed in this paper.

\section{Summary}
We have established a general formalism for ultrafast electron diffractive voltammetry concept, which is applied to investigate the photoinduced charge migration from the substrate to nanostructured interfaces. We show that the surface diffraction and boundary condition need to be accounted for correctly formulating the ultrafast voltammetry based on Coulomb refraction-induced diffracted beam dynamics. We identify that the voltammetry appears on the surface, subsurface, and vacuum levels, associated with interfacial charge transfer, carrier diffusion, and photoemission respectively, under intense laser irradiation.  From quantitative shadow imaging techniques performed at the same condition as voltammetry and N-particle simulations, we are able to assess the voltammetry contribution associated with photoemission, and quantitatively deduce the surface charging dynamics from the overall voltammetry. We find that the photoemission impacts the voltammetry most in the long time, whereas the interfacial charge dynamics dominates the voltammetry on the ultrafast (0-100 ps) timescale.
Miraculously, we are able to extend the diffractive voltammetry methodology to investigate molecular charge transport process under a strong field induced by laser at a gold nanopaticle/molecule/Silicon interface.   At low field circumstance, we obtain similar molecular resistivity as the steady-state measurement, whereas at high fields we observe a molecular dielectric breakdown resembling a field-induced insulator-to-metal transition. The spurious photoemission effect can be suppressed by using a low energy or less intense laser pulse as the high-energy tail of the excited spectrum can be largely cut off under these conditions.  The future applications of this methodology lie in more definitive, site-selected voltammetry studies on nanostructured and heterogeneous interfaces, which can be enabled by the development of nanometer scale high-brightness ultrafast electron beam system for ultrafast electron microscope, which already starts to take shape.\cite{ZewailScience2010,Kim2008,Tao2012}

\section{Acknowledgment}
We acknowledge R. Worhatch, R.K. Raman, Y. Murooka for earlier contribution to nanoparticle work, and valuable exchanges with A.H. Zewail, J.C.H. Spence, and J.M. Zuo. This work is supported under grant DE-FG02-06ER46309 from the Department of Energy. Partial support for R.A. Murdick is under grant 45982-G10 from the Petroleum Research Fund of the American Chemical Society.
\appendix

\section{Formalism of diffracted voltammetry under small angle condition}
\subsection{Slab model}
UEDV formalism can simplified by applying small angle approximation: $\theta_i$ \& $\theta_o \ll 1$, and
$\chi \ll 1$.  Under these conditions, Eq.~(\ref{ECons6}) is reduced to:
\noindent
\begin{widetext}
\begin{equation}
D = \frac{[\theta_o+(1-a)\theta_i]-[(1-a)\sqrt{\theta_i^2+\chi}]}{1+[\theta_o
+(1-a)\theta_i] [(1-a)\sqrt{\theta_i^2+\chi}]}
\sim [\theta_o+(1-a)\theta_i]
-[(1-a)\sqrt{\theta_i^2+\chi]}. \label{AEqn1}
\end{equation}
\end{widetext}
\noindent Also, $\Delta_B$ is modified as
\noindent
\begin{equation}
(\Delta_B+\theta_o)^2 \sim {D^2 - \chi}. \label{AEqn2}
\end{equation}
\noindent By combining Eq.~(\ref{AEqn1}) \& Eq.~(\ref{AEqn2}), we can get the following equation:
\noindent
\begin{widetext}
\begin{equation}
(\Delta_B+\theta_o)^2 = [\theta_o+(1-a)\theta_i]^2 + (1-a)^2\theta_i^2
+ a(a-2)\chi -2(1-a)[\theta_o+(1-a)\theta_i]\sqrt{\theta_i^2+\chi}. \label{AEqn3}
\end{equation}
\noindent Solving Eq.~(\ref{AEqn3}) for $\chi$, we arrive at the following relationship:
\noindent
\begin{equation}
\chi=\frac{B}{2a^2(a-2)^2} \pm \sqrt{ [\frac{B}{2a^2(a-2)^2}]^2-\frac{C}{a^2(a-2)^2}}, \label{AEqn4}
\end{equation}
\noindent where
\noindent
\begin{equation}
B=4(a-1)^2(\theta_i^2+\theta_o^2)+4(a^2-2a+2)(1-a)\theta_i \theta_o+2a(a-2)
(\Delta_B^2+2\Delta_B\theta_o), \label{AEqn5}
\end{equation}
\noindent
\begin{equation}
C=[(\theta_o+(1-a)\theta_i)^2-(\Delta_B +\theta_o)^2]^2+(1-a)^2\theta_i^2
[(1-a)^2\theta_i^2-2(\theta_o+(1-a)\theta_i)^2]. \label{AEqn6}
\end{equation}
\end{widetext}

Simplification can be made for different $a$ value.
For $a=0$, Eq.~(\ref{AEqn3}) can be simplified as:
\begin{equation}
(\Delta_B+\theta_o)^2 = [\theta_o+\theta_i]^2 + \theta_i^2
-2[\theta_o+\theta_i]\sqrt{\theta_i^2+\chi}, \label{AEqn7}
\end{equation}
\noindent so the inversion can be made analytically:
\begin{equation}
\chi=\frac{(\frac{1}{2}\Delta_B^2+\Delta_B\theta_o-\theta_o\theta_i)^2-\theta_i^2(\Delta_B+\theta_o)^2}
{(\theta_o+\theta_i)^2}. \label{AEqn8}
\end{equation}
\noindent which is equivalent to Eq. (1) in the previous study.\cite{Murdick2008}

If $\Delta_B \ll$ $\theta_i$ \& $\theta_o$, Eq.~(\ref{AEqn8}) can be further linearized:
\begin{equation}
\chi=-2\frac{\theta_o^2\theta_i}
{(\theta_o+\theta_i)^2}\Delta_B. \label{AEqn8B}
\end{equation}

For $a=1$, Eq.~(\ref{AEqn3}) can be simplified as:
\begin{equation}
(\Delta_B+\theta_o)^2=\theta_o^2-\chi, \label{AEqn9}
\end{equation}
\noindent yielding a TSV formalism:
\begin{equation}
\chi=-\Delta_B(\Delta_B+2\theta_o).
\label{TSVa1}
\end{equation}

If $\Delta_B \ll$ $\theta_i$ \& $\theta_o$,
\begin{equation}
\chi=-2\theta_o\Delta_B. \label{AEqn9B}
\end{equation}

For $a=2$, implying $\theta_i=\theta_o$ in RHEED geometry, we can reduce the general formalism to
\begin{equation}
(\Delta_B+\theta_o)^2 = \theta_i^2. \label{AEqn10}
\end{equation}
\noindent Therefore, the surface scattered diffraction change, $\Delta_B$, is independent of $\chi$. In other words, the measured $\Delta_B$ in experiments is not affected by TSV.

\subsection{Beyond slab model}
Small angle approximation can also be made for TSV determination beyond the slab model, formulated in Eqs.~(\ref{EqnDeltaB2}) to (\ref{EqnD2}).
For $\Theta_o(\theta_o,\alpha) \neq \Theta_i(\theta_i,\alpha)$, the surface potential, $V_s$ from $\Delta_B$, can be written for small angles as
\begin{widetext}
\begin{align}
\nonumber \lefteqn{\chi=
\nonumber \left\lbrace \vphantom{\frac{1}{1}} \right.  [(c+2)\Theta_i(\theta_i,\alpha)-c \Theta_o(\theta_o,\alpha)]} \qquad \\
\nonumber \\
\nonumber & & \quad \pm \; \sqrt{[(c+2)\Theta_i(\theta_i,\alpha)-c \Theta_o(\theta_o,\alpha)]^2
-4ab[\Theta_o(\theta_o,\alpha)-\Theta_i(\theta_i,\alpha)]^2
} \left. \vphantom{\frac{1}{1}}\right\rbrace  \\
\nonumber \\
& & \left / \vphantom{\frac{1}{1}} \right. b[\Theta_o(\theta_o,\alpha)-\Theta_i(\theta_i,\alpha)]^2, \qquad \qquad
\label{EqnChiLong}
\end{align}
\end{widetext}
\noindent where
\begin{equation}
a = \frac{
\left(
\theta_o \Delta_B - \theta_i \theta_o +\Delta_B^2/2
\right)^2
- \theta_i^2 \left( \theta_o+\Delta_B\right) ^2
} 
{ 
\left( \theta_i+\theta_o  \right)  ^2
},
\end{equation}
\begin{equation}
b=\frac{1}{\left(   \theta_i + \theta_o     \right)^2 },
\end{equation}
\noindent and
\begin{equation}
c = \frac{
2\theta_i^2+2\theta_i\theta_o - 2 \Delta_B \theta_o - \Delta_B^2
}
{
\left(  \theta_i + \theta_o  \right)^2
}.
\end{equation}

\eject



\begin{thebibliography}{0}
\bibitem{RefarctionNote} The shift in diffraction due to electron wavlength change in the materials, which is on the level of $3\times 10^{-5}$ per 1 volt in photovoltage for 30 keV electron beam, is small compared to refraction shift, which is on the $10^{-2}$ level under the same condition.
\bibitem{PersistenceLengthNote} The persistence length here refers to the length of the crystal in the sample that allow the probing electron to scatter coherently to form diffraction pattern.  The persistence length can be limted by the size of the crystal, the coherence length of the probing electron, or the pentration depth of the probing electron, which ever is the smallest.
\bibitem{FieldPrecision} The electrostatics calculations were performed using the Charged Particle Toolkit software from Field Precision. The geometry was setup according to Fig.\ref{FigTSVCorrSlab}(a \& c).  In both cases, a potential of V$_s$=-5 V was imposed. For nanoparticle simulations, the relative permittivity of the dielectric layer was set to 2.5.  The Si(111) was treated as a grounded metal because of the high carrier density under photoexcitation.  The potential far away from the nanoparticle was set to 0.  Electrons of 30 keV were initialized at different launch angles (1$^{\circ}$ to 5$^{\circ}$) and positions along the height of the dielectric layer (1/3, 1/2, and 2/3). The slab model calculations were also carried out with this software, with varying capacitor slab separations.
\end{thebibliography}
\end{document}